\def\answ{b }
\input harvmac
\input labeldefs.tmp
\writedefs
\overfullrule=0pt

\input epsf

\def\fig#1#2#3{
\xdef#1{\the\figno}
\writedef{#1\leftbracket \the\figno}
\nobreak
\par\begingroup\parindent=0pt\leftskip=1cm\rightskip=1cm\parindent=0pt
\baselineskip=11pt
\midinsert
\centerline{#3}
\vskip 12pt
{\bf Fig. \the\figno:} #2\par
\endinsert\endgroup\par
\goodbreak
\global\advance\figno by1
}
\newwrite\tfile\global\newcount\tabno \global\tabno=1
\def\tab#1#2#3{
\xdef#1{\the\tabno}
\writedef{#1\leftbracket \the\tabno}
\nobreak
\par\begingroup\parindent=0pt\leftskip=1cm\rightskip=1cm\parindent=0pt
\baselineskip=11pt
\midinsert
\centerline{#3}
\vskip 12pt
{\bf Tab. \the\tabno:} #2\par
\endinsert\endgroup\par
\goodbreak
\global\advance\tabno by1
}
\def\der{\partial}
\def\d{{\rm d}}
\def\Gam{{\mit\Gamma}}
\def\Gamt{\tilde{\mit\Gamma}}
\def\Th{Thistlethwaite}
\def\HS{Hubbard--Stratonovitch}
\def\e#1{{\rm e}^{#1}}
\def\E#1{{\rm e}^{\textstyle #1}}%

\def\ommit#1{{}}
%
%
%
\lref\AV{I.Ya. Arefeva and I.V. Volovich, 
{\sl Knots and Matrix Models}, {\it Infinite Dim.
Anal. Quantum Prob.} 1 (1998) 1 ({\tt hep-th/9706146}).}
\lref\BIPZ{E. Br{\'e}zin, C. Itzykson, G. Parisi and J.-B. Zuber, 
{\sl Planar Diagrams}, {\it Commun. Math. Phys.} {\bf 59} (1978) 35--51.}
\lref\BIZ{D. Bessis, C. Itzykson and J.-B. Zuber, 
{\sl Quantum Field Theory Techniques in Graphical Enumeration},
{\it Adv. Appl. Math.} {\bf 1} (1980) 109--157.}
\lref\DFGZJ{P. Di Francesco, P. Ginsparg and J. Zinn-Justin, 
{\sl 2D Gravity and Random Matrices, }{\it Phys. Rep.} {\bf 254} (1995)
1--133.}
\lref\tH{G. 't Hooft, 
{\sl A Planar Diagram Theory for Strong 
Interactions}, {\it Nucl. Phys.} {\bf B 72} (1974) 461--473.}
\lref\HTW{J. Hoste, M. Thistlethwaite and J. Weeks, 
{\sl The First 1,701,936 Knots}, {\it The Mathematical Intelligencer}
{\bf 20} (1998) 33--48.}
\lref\MTh{W.W. Menasco and M.B. \Th, 
{\sl The Tait Flyping Conjecture}, {\it Bull. Amer. Math. Soc.} {\bf 25}
(1991) 403--412; 
{\sl The Classification of Alternating 
Links}, {\it Ann. Math.} {\bf 138} (1993) 113--171.}
\lref\Ro{D. Rolfsen, {\sl Knots and Links}, Publish or Perish, Berkeley 1976.}
\lref\STh{C. Sundberg and M. Thistlethwaite, 
{\sl The rate of Growth of the Number of Prime Alternating Links and 
Tangles}, {\it Pac. J. Math.} {\bf 182} (1998) 329--358.}
\lref\Tutte{W.T. Tutte, {\sl A Census of Planar Maps}, 
{\it Can. J. Math.} {\bf 15} (1963) 249--271.}
\lref\Zv{A. Zvonkin, {\sl Matrix Integrals and Map Enumeration: An Accessible
Introduction},
{\it Math. Comp. Modelling} {\bf 26} (1997) 281--304.}
\lref\KM{V.A.~Kazakov and A.A.~Migdal, {\sl Recent progress in the
theory of non-critical strings}, {\it Nucl. Phys.} {\bf B 311} (1988)
171--190.}
\lref\KP{V.A.~Kazakov and P.~Zinn-Justin, {\sl Two-Matrix Model with
$ABAB$ Interaction}, {\it Nucl. Phys.} {\bf B 546} (1999) 647
({\tt hep-th/9808043}).}
\lref\ZJZ{P.~Zinn-Justin and J.-B.~Zuber, {\sl Matrix Integrals
and the Counting of Tangles and Links},
to appear in the proceedings of the 11th 
International Conference on Formal Power Series and Algebraic 
Combinatorics, Barcelona June 1999
(preprint {\tt math-ph/9904019}).}
\lref\PZJ{P.~Zinn-Justin, {\sl Some Matrix Integrals
related to Knots and Links}, to appear in the proceedings
of the 1999 semester of the MSRI on Random Matrices
(preprint {\tt math-ph/9910010}).}
\lref\PZJb{P.~Zinn-Justin, {\sl The Six-Vertex Model on
Random Lattices}, preprint {\tt cond-mat/9909250}.}
\lref\IK{I.~Kostov, {\sl Exact solution of the Six-Vertex
Model on a Random Lattice}, preprint {\tt hep-th/9911023}.}
\lref\KAUF{L.H.~Kauffman, {\sl Knots and physics},
World Scientific Pub Co (1994).}
\Title{
\vbox{\baselineskip12pt\hbox{RUNHETC-2000-04}\hbox{SPhT 00/007}\hbox{{\tt math-ph/0002020}}}}
{{\vbox {
\vskip-10mm
\centerline{On the Counting of Colored Tangles}
}}}
\medskip
\centerline{P.~Zinn-Justin}\medskip
\centerline{\it New High Energy Theory Center}
\centerline{\it Department of Physics and Astronomy, Rutgers University,} 
\centerline{\it Piscataway, NJ 08854-8019, USA}
\bigskip
\centerline{and}
\medskip
\centerline{J.-B.~Zuber}\medskip
\centerline{\it C.E.A.-Saclay, Service de Physique Th{\'e}orique,}
\centerline{\it F-91191 Gif sur Yvette Cedex, France}

\vskip .2in
\noindent
The connection between matrix integrals and links is
used to define matrix models which
count alternating tangles in which each closed loop 
is weighted with a factor $n$, i.e.\ may be regarded as
decorated with $n$ possible colors. For $n=2$, the corresponding 
matrix integral is that recently solved in the study 
of the random lattice six-vertex model. The generating 
function of alternating 2-color tangles is provided in terms
of elliptic functions, expanded to 16-th order (16 crossings)
and its asymptotic behavior is given.

\Date{02/2000}

\newsec{Introduction}
The problem of counting topologically distinct knots remains
a challenging one: see \HTW\ for a review and for a report 
on recent advances. It was recently noticed \ZJZ\ that combinatorial
methods developed in quantum field theory, namely Feynman diagrams
\ applied to matrix integrals, may provide a new way to count
knots. In particular, the counting of {\it alternating\/} tangles,
which had been achieved in \STh, was reproduced. This counting, 
however, is not capable of discriminating between objects with 
different numbers of connected components: in the usual terminology, 
it gives the number of links rather than that of knots. In the 
present article,
we reexamine this question and show how the introduction of a number $n$
of possible ``colors'' for the knotted loops would solve this question, 
and how this may be formulated in terms of a matrix integral. 
For $n=2$, this integral is equivalent to one recently studied 
in detail and computed in the framework of the random lattice
6-vertex model \refs{\PZJb{--}\IK}. We thus  carry out the explicit 
counting of alternating 2-color tangles: their generating 
function is the solution of coupled equations involving elliptic
functions. We are able to give the 13 first terms of its
expansion and its asymptotic behavior. We conclude with more 
conjectural considerations on the number of (2-color alternating)
links.

\newsec{A matrix model for colored links}
We want to consider a model which describes alternating links in which each
of the intertwined loops can have $n$ different colors. This can be
achieved via a large 
$N \times N$ matrix integral \ZJZ. As a first attempt,
let us consider the following integral
\eqn\mmm{
Z^{(N)}(n,g)=\int\! \prod_{a=1}^n \d M_a
\, \E{N\,\tr\left(-{1\over 2} \sum_{a=1}^n M_a^2+{g\over 4} \sum_{a,b=1}^n
M_a M_b M_a M_b\right)}}
over $N \times N$ hermitean matrices, 
and the corresponding ``free energy''
\eqn\mmmb{F(n,g)=\lim_{N\to\infty}{\log Z^{(N)}(n,g)\over N^2}\ .}
Such an integral has a $g$-series expansion (``perturbative 
expansion'') which admits a graphical representation in terms 
of Feynman diagrams. In the large $N$ limit, only planar diagrams 
survive: see for example \BIZ\ for a review. 
The integral \mmm\ 
has an $O(n)$-invariance where the $M_a$ 
form the vector representation of 
$O(n)$. The Feynman diagram expansion of $F(n,g)$ generates
planar diagrams with four-legged
vertices and colored edges such that colors cross each other
at each vertex (Fig.~\feya).

To each planar diagram we can associate
an alternating link diagram (see e.g.\ \KAUF, page 21) 
by following colored loops as they cross other loops
and choosing alternatingly under- and over-crossings. This can
be carried out in a consistent way throughout the whole diagram.
\fig\feya{A planar Feynman diagram of \mmmb\ and the corresponding
alternating link diagram.}{\epsfbox
{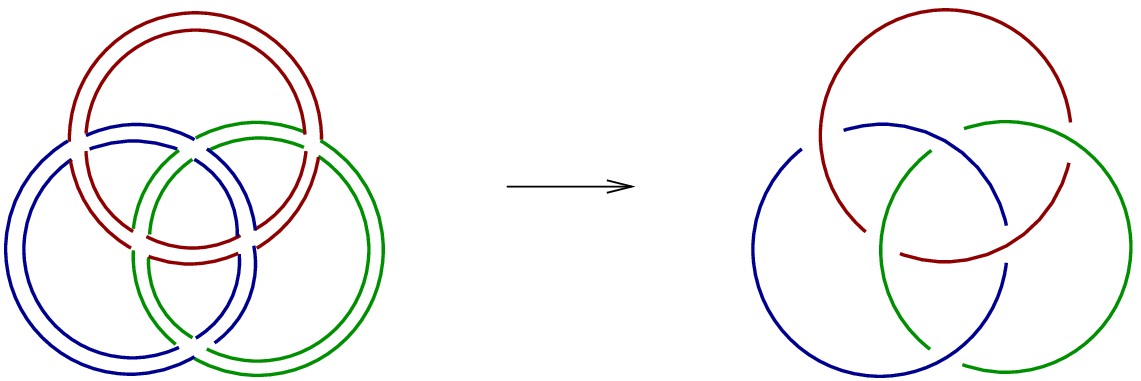}}

We have thus generated alternating link diagrams with $n$ colors.
Diagrams with different numbers of connected components can
now be distinguished by the $n$ dependence of their weight: indeed,
to a diagram with $k$ connected components is associated
a factor $n^k$. We can therefore write
\eqn\mmmc{F(n,g)=\sum_{k=1}^\infty F_k(g) n^k}
where $F_k(g)$ is the sum over alternating link diagrams with
exactly $k$ intertwined loops. Note that if one can
define $F(n,g)$ for non-integer $n$ and in particular in
a neighborhood of $0$, 
one has access to the individual contributions $F_k(g)$ since
they form the small $n$ expansion of $F(n,g)$.
As usual in $O(n)$ vector models, one can perform (at least
formally) 
an analytic continuation in $n$ by a \HS\ transformation.
Unfortunately such a transformation breaks planarity of the
diagrams and is not suitable for our purposes. However, another \HS\ 
transformation exists for the particular values $n=\pm 2$ (besides
$n=1$, of course) which preserves planarity, as will be explained
later.

We can also define correlation functions in the model. There is only
one 2-point function,
\eqn\twopoint{G(g)=\lim_{N\to\infty}\left< {1\over N}\tr M_a^2\right>}
where $a$ is fixed, $1\le a\le n$. 
Here the brackets $\left<\cdot\right>$ refer to the normalized 
average with the exponential weight of \mmm.
Due to $O(n)$-invariance, there
are {\it two\/} independent 4-point functions. We consider connected
4-point functions only; we choose
\eqna\fourpoint
$$\eqalignno{
\Gam_1(g)&=\lim_{N\to\infty}\left< {1\over N}\tr (M_a M_b)^2\right>_{\rm c}&\fourpoint{\rm a}\cr
\Gam_2(g)&=\lim_{N\to\infty}\left< {1\over N}\tr (M_a^2 M_b^2)\right>_{\rm c}&\fourpoint{\rm b}\cr
}$$
where $a$ and $b$ are fixed and distinct. They have the following
interpretation: $\Gam_i(g)$
is the generating function of the
number of colored alternating tangle diagrams of type $i$;
a diagram with four external legs is of type $1$ if the external
strings come in and out in diagonally opposite corners; it
is of type $2$ if the external strings come in and out in the
same upper/lower half-plane; see Fig.~\types. Note that any diagram with four
external legs is either
of type $1$, or of type $2$, or the image of a type $2$ diagram
under a $\pi/2$ rotation.
\fig\types{The two types of tangle 
diagrams.}{\epsfxsize=11cm\epsfbox{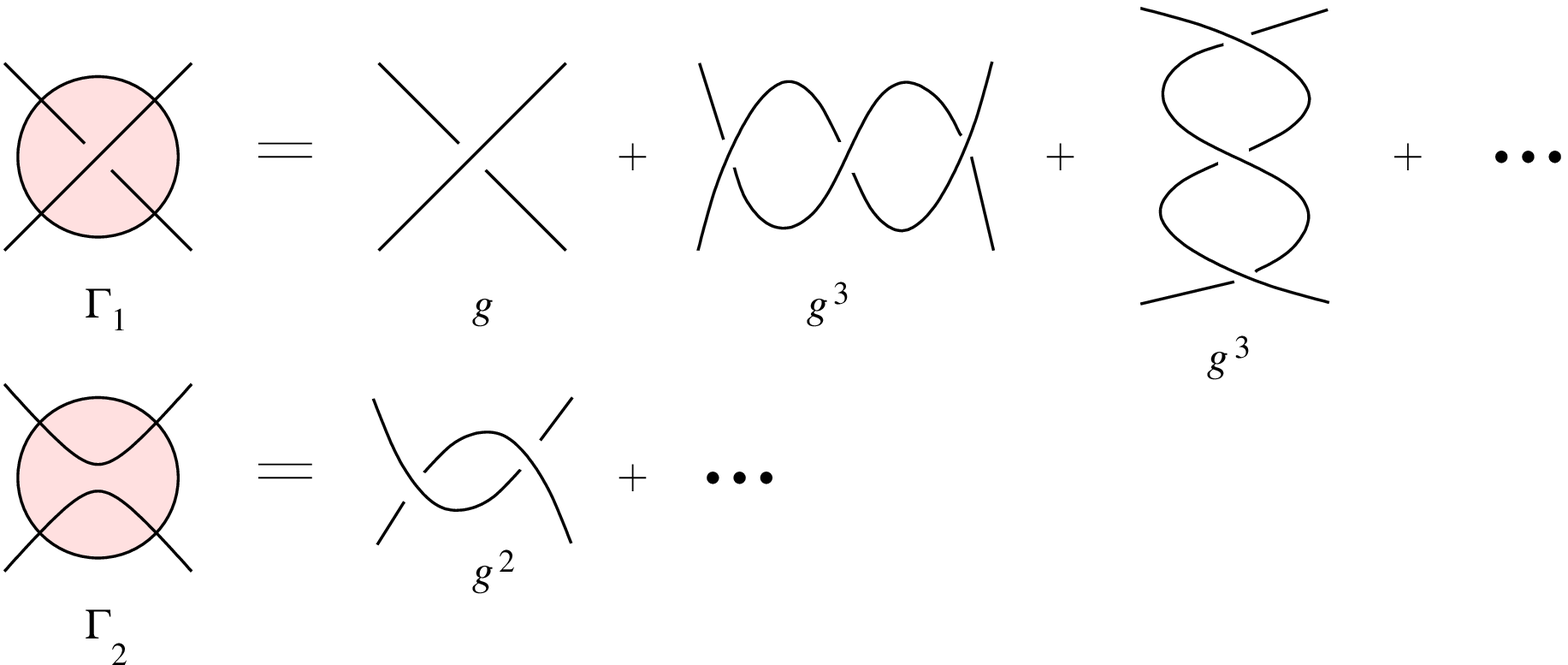}}

We must now proceed as in \ZJZ. Our goal is 1) to count only diagrams
which are prime and reduced, and 2) to count as a single contribution
diagrams which correspond to the same link.
First it is necessary to
take care of non-prime and non-reduced diagrams. This is achieved
by introducing an extra parameter $t$ in the action:
\eqn\mmm{
Z^{(N)}(n,t,g)=\int\! \prod_{a=1}^n \d M_a
\, \E{N\,\tr\left(-{t\over 2} \sum_{a=1}^n M_a^2+{g\over 4} \sum_{a,b=1}^n
M_a M_b M_a M_b\right)}\ ,}
and choosing $t$ as a function of $g$ in such a way that
\eqn\propone{G(t(g),g)=1\ .}
We have the obvious scaling property $G(t,g)={1\over t} G(1,g/t^2)$,
which means that given the two-point function of the
original model $G(1,g)\equiv G(g)$, $t(g)$ is the solution
of the equation:
\eqn\propeqn{t(g)=G(g/t^2(g))\ .}
We have similar scaling properties for the higher
correlation functions; in particular, $\Gam_i(t,g)={1\over t^2}
\Gam_i(1,g/t^2)$.

\fig\flype{The flype of a tangle}{\epsfxsize=6cm\epsfbox{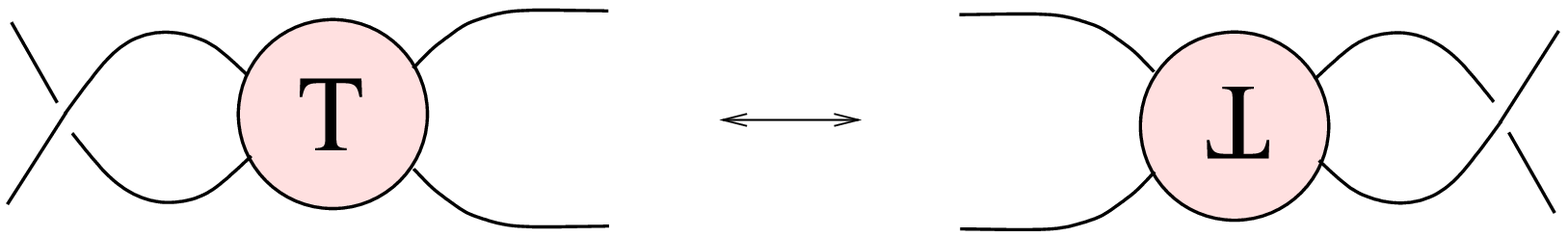}}
The next step is to remove the over-counting of links
due to the fact that several diagrams can correspond to a single
link. According to the Tait flyping conjecture (proven in \MTh),
two reduced alternating diagrams are equivalent if and only if
they are related by a sequence of flypes (Fig.~\flype). In order to take
into account the flyping equivalence,
we must now consider, as advocated in \PZJ, 
the most general $O(n)$-invariant model with quartic interaction
\eqn\mmmgen{
Z^{(N)}(n,t,g_1,g_2)=\int\! \prod_{a=1}^n \d M_a
\, \E{N\,\tr\left(-{t\over 2} \sum_{a=1}^n M_a^2
+{g_1\over 4} \sum_{a,b=1}^n (M_a M_b)^2
+{g_2\over 2} \sum_{a,b=1}^n M_a^2 M_b^2
\right)}\ .}
There is a new type of vertex which allows loops to ``avoid'' each
other. The appearance of two types of vertices can be understood
as follows: the unwanted (i.e.\ overcounted) 
tangle diagrams due to the flyping equivalence can be
of either type 1 or type 2, and we must introduce ``counterterms''
of both types to cancel them. To pursue the analogy with
renormalization, we can rephrase this by saying that
it is only the ``renormalized'' coupling constants
which must be of the form $(g,0)$, but in order to reach this
point we must consider more general ``bare'' coupling
constants $(g_1,g_2)$.

We impose again the condition
\eqn\proponeb{G(t(g_1,g_2),g_1,g_2)=1\ .}
{}From now on $t$ will be assumed to be fixed by \proponeb, and we shall
be concerned with the correlation functions $\Gam_i\equiv
\Gam_i(t(g_1,g_2),g_1,g_2)$, which are the generating functions of
diagrams of type $i$ with four external legs and two types of vertices 
weighted by $g_1$ and $g_2$ respectively.

As in \ZJZ, we make use of the concepts of 
two-particle irreducibility. 
Following the language of field theory,
we recall that a four-legged diagram
is {\it two-particle-irreducible} (2PI)
if cutting any two distinct propagators leaves it connected;
otherwise it is {\it two-particle-reducible} (2PR).
Also, we shall consider
skeleton diagrams, which must be ``dressed'' to recover
ordinary diagrams. This will be implicit in the following
and we refer the reader to \ZJZ\ for details.

Let us define $D_i$ (resp.\ $H_i$, $V_i$) to be the generating functions
of 2PI (resp.\ 2PI in the horizontal, vertical channel)
four-legged diagrams of type $i$.
Note that $H_1=V_1$, but $H_2\ne V_2$, 
since the defining property of diagrams of type $2$
is not invariant by rotation of $\pi/2$.

By decomposing a general diagram $\Gam_1$ according to the number
of times it is reducible in the {\it horizontal\/} channel (Fig.~\horiz),
we find the following formulae:
\eqna\dec
$$\eqalignno{
\Gam_1&={1\over 2}\left( {H_2+H_1\over 1-(H_2+H_1)}
-{H_2-H_1\over 1-(H_2-H_1)}\right)&\dec{\rm a}\cr
\Gam_2&={1\over 2}\left( {H_2+H_1\over 1-(H_2+H_1)}
+{H_2-H_1\over 1-(H_2-H_1)}\right)\ .&\dec{\rm b}\cr
}$$
\fig\horiz{Decomposition of a tangle diagram in the horizontal channel.}
{\epsfxsize=12cm\epsfbox
{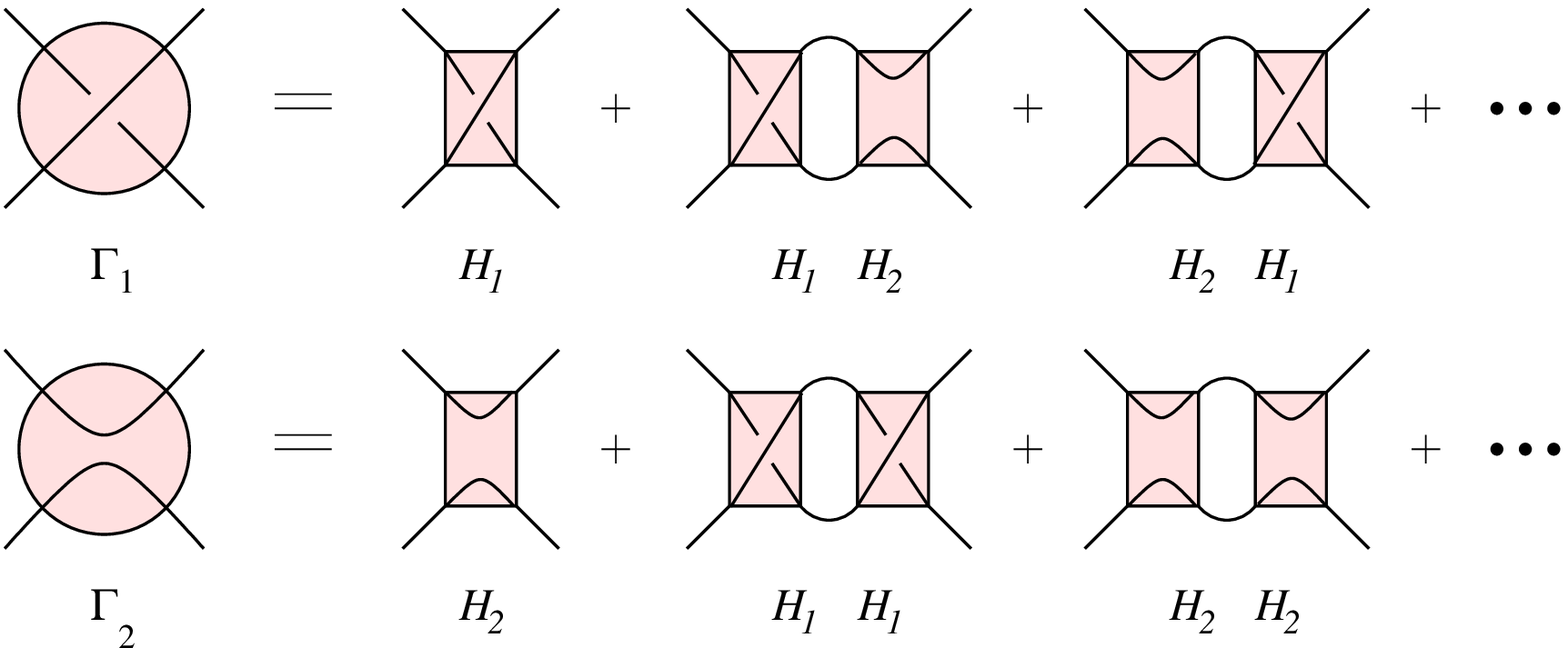}}
This can be simplified by introducing the combinations
$\Gam_\pm =\Gam_2\pm \Gam_1$ and similarly for $H_\pm $. We find:
$$\Gam_\pm ={H_\pm \over 1-H_\pm }\ .\eqno \dec{'}$$
Note that these equations are independent of $n$.

It is not as obvious how 
to decompose diagrams of type $2$ using
the {\it vertical\/} channel. It is simpler to proceed as follows:
define $\Gam_0$ to be generating function of diagrams with
four external legs such that the color of the two left outgoing
strings is free, whereas the color of the two right outgoing
strings is fixed and equal; and similarly $H_0$, $V_0$, $D_0$.
We have the formulae
\eqna\defzero
$$\eqalignno{
\Gam_0&=(n+1)\Gam_2+\Gam_1&\defzero{\rm a}\cr
H_0&=H_2+n V_2+H_1 \ . &\defzero{\rm b}\cr
}$$
We can now proceed to decompose $\Gam_0$ in the horizontal channel.
It is easy to convince oneself that the simple formula
\eqn\decc{\Gam_0={H_0\over 1-H_0}}
holds.

The three equations \dec{} and \decc\ 
determine $H_1=V_1$, $H_2$ and $V_2$ as
functions of $\Gam_1$ and $\Gam_2$.\foot{If one sets $n=0$,
two equations become identical and one has to include
the derivative of \decc\ with respect to $n$:
$\Gam_2=V_2/(1-(H_1+H_2))^2$.}
More explicitly, one should invert them to 
\eqn\decd{H_{0,\pm }={\Gam_{0,\pm }\over 1+\Gam_{0,\pm }}}
and take appropriate linear combinations.
Furthermore, noting that any diagram with four external legs
and four-legged vertices must be irreducible in one of the two
channels, we have
\eqn\dece{D_i=H_i+V_i-\Gam_i\ ,}
which means that we have managed to express the $D_i$ in
terms of the $\Gam_i$.

The flype equivalence does not affect 2PI skeleton diagrams; this
means in practive that
the expression of $D'_i=D_i-g_i$ as a function of the $\Gam_i$
is left unchanged by the removal of the overcounting of flype
equivalent diagrams. Note that the $D_i[\Gam_1,\Gam_2]$
have a simple expression given by Eqs. \decd--\dece, but
the $g_i[\Gam_1,\Gam_2]$
are non-trivial functions which can only be obtained by
actually solving the matrix model \mmmgen\ (or some
equivalent procedure).
For example, perturbatively we find: (Fig.~\perturb)
\eqna\pertb
$$\eqalignno{
D'_1&=n\Gam_1^5+8\Gam_1^4\Gam_2+(4n+4)\Gam_1^3\Gam_2^2
+24\Gam_1^2\Gam_2^3+16\Gam_1^6\Gam_2+O(g^9)&\pertb{\rm a}\cr
D'_2&=n\Gam_1^4\Gam_2+2\Gam_1^7+16\Gam_1^3\Gam_2^2+(20+8n)\Gam_1^2\Gam_2^3
+(14+6n)\Gam_1^6\Gam_2+3\Gam_1^8+O(g^9)\cr&&\pertb{\rm b}\cr
}$$
where the order of truncation is dictated by the rule:\foot{This rule
can be justified as follows: at leading order, the bare
coupling constants can be replaced with their renormalized values:
$g_1=g$, $g_2=0$. One then finds $\Gam_1=g+O(g^2)$, $\Gam_2=g^2+O(g^3)$.}
$\Gam_1\sim g$, $\Gam_2\sim g^2$.
\fig\perturb{First terms in the perturbative
expansion of $D'_1$ and $D'_2$.}{\epsfxsize=12cm\epsfbox{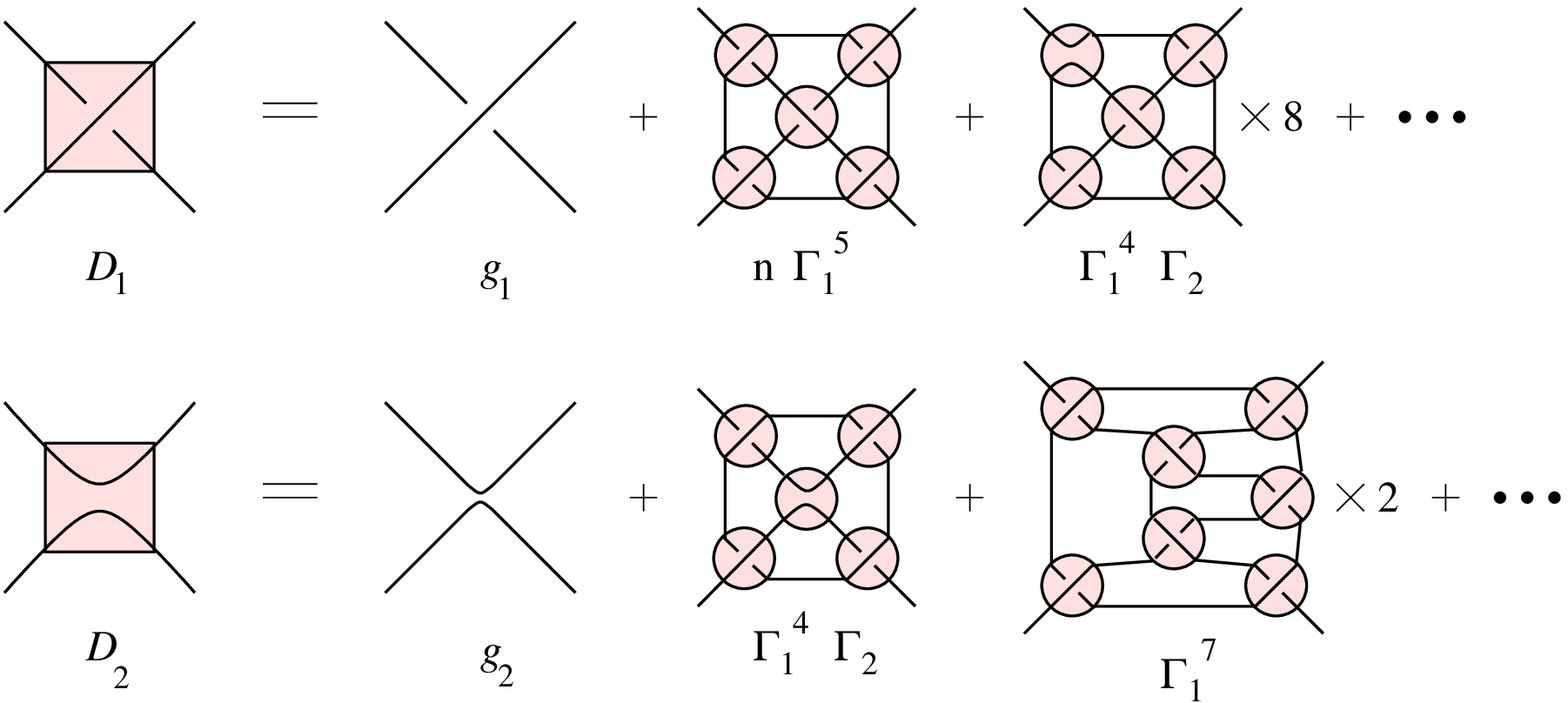}}

The data $D'_i[\Gam_1,\Gam_2]$ is all we need from the
generalized matrix model \mmmgen, and we can now come back to the problem
of tangle diagrams, with a single coupling constant $g$ weighting
a simple crossing.

We want to redo the decompositions of a general diagram in the
horizontal channel, but this time taking into account the flyping
equivalence. This forces us to distinguish between simple crossings
and other diagrams, that is to use the primed objects defined
by $\Gamt_1'=\Gamt_1-g$, $\tilde{H}_1'=\tilde{H}_1-g$,
$\tilde{V}'_1=\tilde{V}_1-g$. 
Here, as in \ZJZ, the tilde on one of the symbols $\Gamma$, $H$ 
or $V$  denotes generating functions 
of flype equivalence classes of 2PI skeleton diagrams.
We find:
\eqna\decf
$$\eqalignno{
\Gamt_1'&=g \Gamt_2+{1\over 2}\left( {\tilde{H}_2+\tilde{H}'_1\over 1-(\tilde{H}_2+\tilde{H}'_1)}
-{\tilde{H}_2-\tilde{H}'_1\over
 1-(\tilde{H}_2-\tilde{H}'_1)}\right)&\decf{\rm a}\cr
\Gamt_2'&=g \Gamt_1+{1\over 2}\left( {\tilde{H}_2+\tilde{H}'_1\over 1-(\tilde{H}_2+\tilde{H}'_1)}
+{\tilde{H}_2-\tilde{H}'_1\over
1-(\tilde{H}_2-\tilde{H}'_1)}\right)\ .&\decf{\rm b}\cr
}$$
Introducing $\tilde{H}'_\pm =\tilde{H}'_2\pm \tilde{H}_1$, we rewrite this
\eqn\decg{(1\mp g)\Gamt_\pm 
=\pm g+{\tilde{H}'_\pm \over 1-
\tilde{H}'_\pm }\ .}
Similarly, if $\tilde{H}'_0=\tilde{H}_0-g$, we find
\eqn\dech{
(1-g)\Gamt_0=g+{\tilde{H}'_0\over 1-\tilde{H}'_0}\ .}
We invert these relations to
\eqna\deci
$$\eqalignno{
\tilde{H}'_\pm &={(1\mp g)\Gamt_\mp\mp g\over 1+(1\mp g)\Gamt_\mp\mp g}
&\deci{\rm a}\cr
\tilde{H}'_0&={(1-g)\Gamt_0-g\over 1+(1-g)\Gamt_0-g}
&\deci{\rm b}\cr
}$$
and express the 2PI functions
\eqna\decj
$$\eqalignno{
\tilde{D}'_1&=\tilde{H}'_1+\tilde{V}'_1-\Gamt'_1=
\tilde{H}'_+ - \tilde{H}'_- - \Gamt_1 - g
&\decj{\rm a}\cr
\tilde{D}'_2&=\tilde{H}'_2+\tilde{V}'_2-\Gamt'_2=
{1\over2}(\tilde{H}'_+ + \tilde{H}'_-) + {1\over n}(\tilde{H}'_0
-\tilde{H}'_+) - \Gamt_2 \ .
&\decj{\rm b}\cr
}$$

Finally, the generating functions $\Gamt_i(g)$ are given by
the implicit equations
\eqn\finaleq{
\tilde{D}'_i(g)=D'_i[\Gamt_1(g),\Gamt_2(g)]
}
where $D'_i[\Gam_1,\Gam_2]$ comes from the solution of the matrix
model, and $\tilde{D}'_i(g)$ is given by Eqs.~\deci{}--\decj{}\ 
with $\Gamt_i=\Gamt_i(g)$.

We can solve the implicit equation \finaleq\ perturbatively using
\pertb{}; the result is summarized in Tab.~\tableaun.
\tab\tableaun{Table of the number of prime alternating
tangles up to $8$ crossings. The
power of $n$ indicates the number of closed loops, i.e.\ the
number of connected components besides
the two outgoing strings.}{\vbox{\offinterlineskip
\halign{\strut\hfil$#$\hfil\quad&\vrule#&&\quad$#$\hfil\crcr
&&\hfil g&\hfil g^2&\hfil g^3&\hfil g^4&\hfil g^5&\hfil g^6&\hfil
g^7&\hfil g^8\cr
\omit&height2pt\cr
\noalign{\hrule}
\omit&height2pt\cr
\Gamt_1&&1&&2&2&6+3n&30+2n&62+40n+2n^2&382+106n+2n^2\cr
\Gamt_2&&&1&1&3+n&9+n&21+11n+n^2&101+32n+n^2&346+153n+24n^2+n^3\cr
}}}
These results are compatible with the exact solution at $n=1$ \STh,
and they will give us a non-trivial check of our new solution at
$n=2$.


\newsec{The $n=2$ case: a model of oriented links}
Let us now carry out explicitly the procedure outlined in the previous
section, for a value of $n$ corresponding to a solved matrix model.
The case $n=1$ has already been considered in \ZJZ; let us now set
$n=2$.
The partition function
\eqn\twomm{
Z=\int \d M_1\d M_2
\, \E{N\,\tr\left(-{t\over 2} (M_1^2+M_2^2)
+{g_1+2g_2\over 4}(M_1^4+M_2^4)
+{g_1\over 2} (M_1 M_2)^2
+g_2 M_1^2 M_2^2
\right)}}
is conveniently rewritten
in terms of a complex matrix $X=\sqrt{t\over2}\left(M_1+iM_2\right)$:
\eqn\sixv{
Z=\int \d X\d X^\dagger \,\E{N\,\tr\left(-XX^\dagger+b_0 X^2X^\dagger{}^2
+{1\over 2}c_0(XX^\dagger)^2\right)}}
where we have
absorbed $t$ in the coupling constants: $b_0=b/t^2$, $c_0=c/t^2$,
with $b=g_1+g_2$ and $c=2g_2$.
It is clear that the partition function has been left unchanged
by the transformation, but the individual diagrams are different.
The new Feynman rules are depicted on Fig.~\feyb\ a).
If we simply set $b=g$, $c=0$ this model describes oriented links, 
each closed loop  having now two possible orientations instead of two
colors.
\fig\feyb{Feynman rules corresponding to a) \sixv\ and 
b) \sixvb.}{\epsfbox{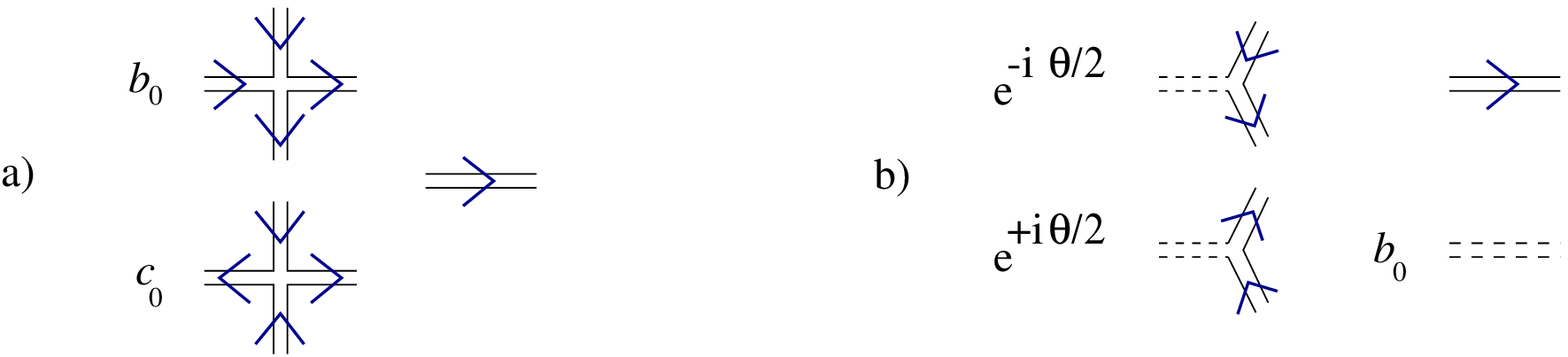}}

We recognize in \sixv\ the partition function of
the {\it six-vertex model\/} on random dynamical lattices, which
has recently been solved \refs{\PZJb,\IK}.
In the 6 vertex formulation it is particularly clear that there is
a \HS\ transformation which preserves planarity;\foot{The
same remark applies to the corresponding fermionic model,
i.e.\ $n=-2$.}
introducing the notation $c=2b\cos\theta$, with $\theta\in[0,\pi[$
in the regime of interest to us, we have
\eqn\sixvb{Z=\int \d A \d X\d X^\dagger \E{N\tr\left(
-XX^\dagger-{1\over 2b_0}A^2+A(XX^\dagger \e{i\theta/2}+
X^\dagger X \e{-i\theta/2})\right)}
}
where $A$ is a hermitean matrix.
(Here and in the following integral \sixvc, we omit an overall constant
factor, irrelevant in the computation of correlation functions.)
We have rewritten our model as a model of oriented loops which are {\it not\/}
the intersecting loops we started from, since they can {\it only\/}
avoid each other, see Fig.~\feyb\ b). This may seem unnatural, but is what
allows to solve exactly the model, since we can now integrate over
the original matrix $X$, 
shift the matrix $A$:
\eqn\sixvc{Z=\int \d A \det{}^{-1}\left(\e{i\theta/2}\otimes A+
A\otimes\e{-i\theta/2}\right)
\E{-N{1\over2b_0}\tr \big(A-{1\over2\cos(\theta/2)}\big)^2}
}
and do a saddle point analysis of the eigenvalues of $A$.
This is the basis of the analytic solutions \refs{\PZJb,\IK} of the model.

It is natural to redefine the two independent 4-point functions
to be
\eqna\fourpointb
$$\eqalignno{
\Gam_b&=\left< {1\over N} \tr (X^2X^\dagger{}^2)\right>_{\rm c}
&\fourpointb{\rm a}\cr
\Gam_c&=\left< {1\over N} \tr (XX^\dagger)^2\right>_{\rm c}
&\fourpointb{\rm b}\cr
}$$
i.e.\ they are characterized by the position of the
ingoing/outgoing arrows on the external legs.
They are related to the 4-point functions defined earlier by:
$\Gam_b=\Gam_1+\Gam_2$, $\Gam_c=2\Gam_2$, as is clear diagrammatically.

Let us summarize the combinatorial relations of the previous section
in the case $n=2$.
First, the relations which determine
the corresponding 2PI correlation functions $D_b$ and $D_c$
can be obtained
either by direct diagrammatic arguments or by setting $n=2$ in the
general formulae \dec{}--\decc. We find:
\eqna\dectwo
$$\eqalignno{
\Gam_b&={H_b\over 1-H_b}&\dectwo{\rm a}\cr
\Gam_c\pm \Gam_b&={V_c\pm V_b\over 1-(V_c\pm V_b)}
&\dectwo{\rm b}\cr
D'_b&=H_b+V_b-\Gam_b-b&\dectwo{\rm c}\cr
D'_c&=H_c+V_c-\Gam_c-c \ .&\dectwo{\rm d}\cr
}$$
Similarly, the modified relations which take into account the
flyping equivalence read: (we
have reintroduced the renormalized coupling constant
$g$ corresponding to a simple crossing, and $\tilde{H}'_b=\tilde{H}_b-g$, etc)
\eqna\dectwob
$$\eqalignno{
\Gamt_b&=g(1+\Gamt_b)+{\tilde{H}'_b\over 1-\tilde{H}'_b}
&\dectwob{\rm a}\cr
(1\mp g)(\Gamt_c\pm \Gamt_b)&= \pm g
+{\tilde{V}_c\pm \tilde{V}'_b\over 1-(\tilde{V}_c\pm \tilde{V}'_b)}
&\dectwob{\rm b}\cr
\tilde{D}'_b&=\tilde{H}_b+\tilde{V}_b-\Gamt_b-g&\dectwob{\rm c}\cr
\tilde{D}'_c&=\tilde{H}_c+\tilde{V}_c-\Gamt_c \ .&\dectwob{\rm d}\cr
}$$

We shall now use the solution of the matrix model \sixv\ \refs{\PZJb,\IK}
to show how to extract the functions $\Gamt$ perturbatively
at an arbitrary order, and to find their large order behavior.


\subsec{Perturbative expansion of the off-critical solution.}
In \IK, the large $N$ saddle point density of eigenvalues of $A$
is explicitly constructed in terms of elliptic functions.
More precisely if we define the resolvent 
\eqn\res{
W(a)=\lim_{N\to\infty} \left<{1\over N}\tr{a\over a-A}\right>}
where the average is with respect to the measure in \sixvc,
and 
\eqn\resb{J(a)=i\left[W\big(i\,a\,\e{-i\theta/2}\big)
-W\big(-i\,a\,\e{i\theta/2}\big)\right]+{1\over2\sin\theta\, b_0} a^2
-{1\over 4\cos^2(\theta/2)\,b_0} a
}
then we have the following expression for $J$ using
an elliptic parametrization $u$:
\eqna\ellipt
$$\eqalignno{
J&=A+B{1\over{\rm sn}^2(u-u_\infty)}&\ellipt{\rm b}\cr
a&=a_0 {H(u_\infty+u)\over H(u_\infty-u)}&\ellipt{\rm a}\cr
}$$
where $H$ is the Jacobi theta function and $A$, $B$, $u_\infty$
$a_0$ are constants which depend on $b_0$ and $\theta$.

If we consider the small $b$, $c$ perturbative expansion of
the correlation functions of the model, we are in the region where
the elliptic nome $q$ is close to zero, and the elliptic functions
can be expressed to a given order in $q$ in terms of trigonometric
functions.
We can therefore write every quantity as a power series in $q$ with
coefficients dependent on $\theta$. 
{}From the $1/a$ term in the $a\to\infty$ expansion
of $J(a)$ we can extract
$W_1=\lim_{N\to\infty}\left<{1\over
N}\tr A\right>$, and from there the two-point function
using the formula
\eqn\perta{
G={1\over b_0}\left({1\over2\cos(\theta/2)}-W_1\right)\ .}
We can then go back to the free energy by integrating once (at fixed $\theta$)
\eqn\pertb{
F=\int^{b_0} \d b_0 {G-1\over 2b_0}}
and differentiate again to get the 4-point functions:
\eqn\pertc{
H={\der\over\der\theta}F_{\left| b_0\ {\rm fixed}\right.}\ .}
The rescaling which allows to set $G=1$ simply amounts to
writing that the coupling constant $b$ is
\eqn\pertd{b=b_0\, G^2\ .}
Then we have
\eqna\perte
$$\eqalignno{
\Gam_b&={(G-1)/2+H\cot\theta\over b}-1&\perte{\rm a}\cr
\Gam_c&=-{H\over b\sin\theta}-2\ .&\perte{\rm b}\cr
}$$
We must apply formulae \dectwo{}\ to evaluate
$D'_b$ and $D'_c$. Finally, we must consider $q$ and $\theta$
as power series in the renormalized coupling constant $g$ and solve
perturbatively in $g$ the equations \dectwob{}.

This can be programmed
on a computer using for example Mathematica$^{\rm TM}$.
At the first non-trivial orders we find:
$$\eqalign{
b_0&=q^2-6(1+2\cos\theta)q^4+O(q^6)\cr
G&=1+2(1+2\cos\theta)q^2+O(q^4)\cr
\Gam_b&=q^2-q^4+O(q^6)\cr
\Gam_c&=2\cos\theta\, q^2+2(1-\cos\theta)q^4+O(q^6)\cr
D'_b&=(6+12\cos\theta+4\cos(2\theta)+4\cos(3\theta))q^{10}+O(q^{12})\cr
D'_c&=(8+24\cos\theta+8\cos(2\theta)+10\cos(3\theta)+2\cos(5\theta))q^{10}+O(q^{12})\ .\cr
}$$
The next step is to solve perturbatively
the Eqs.~\dectwob{}. We can theoretically proceed to an
arbitrary order in $q$ and
therefore in $g$. In Tab.2 
we show the results obtained 
after eight hours on a Sun work-station.
\ommit{{\tab\tableautwo{Table of the number of oriented prime
alternating tangles up to $13$
crossings, and related quantities. $\Delta\theta\equiv\theta-\pi/2$.}
{\vbox{\offinterlineskip
\halign{\strut\hfil$#$\hfil\quad&\vrule#\enskip&&\enskip\hfil$#$\hfil\crcr
&&g&g^2&g^3&g^4&g^5&g^6&g^7&g^8&g^9&g^{10}&g^{11}&g^{12}&g^{13}\cr
\omit&height2pt\cr
\noalign{\hrule}
\omit&height2pt\cr
q^2 && 1&2&7&29&137&679&3515&18677&101463&560062&3132639
&17708417&100998567\cr
\Delta\theta&&&1&1&4&13&{319\over6}&{437\over2}&{1941\over2}&
{13424\over3}&{858263\over40} &{844871 \over 8}&{6386963\over12}&\ldots\cr
b && 1&&-1&-3&-9&-27&-103&-411&-1838&-8484&-41000&-202822&-1027954\cr
c && &&-2&-2&-6&-18&-74&-314&-1420&-6696&-32592&-162728&-828344\cr
g_1&& 1&&&-2&-6&-18&-66&-254&-1128&-5136&-24704&-121458&-613782\cr
g_2&& &&-1&-1&-3&-9&-37&-157&-710&-3348&-16296&-81364&-414172\cr
\Gamt_b&& 1&1&3&7&23&81&319&1358&6132&28916&140852&704020&3592394\cr
\Gamt_c&& &2&2&10&22&94&338&1512&6700&31944&155200&778168&3972088\cr
\omit&height2pt\cr
\noalign{\hrule}
\omit&height2pt\cr
\Gamt_1&&1&&2&2&12&34&200&602&2782&12944&63252&314936&1606350\cr
\Gamt_2&& &1&1&5&11&47&119&756&3350&15972&77600&389084&1986044\cr
}}}
}}
As a check, note that $g_1=(1-\cos\theta)b=g+O(g^4)$ and 
$g_2=\cos\theta\, b=-g^3+O(g^4)$ which is consistent
with the fact that
at order $3$ there is exactly one diagram of type 2
(up to a $\pi/2$ rotation) which is overcounted (see Fig.~\first).
Also, these data are compatible with those of Tab.~\tableaun.
\fig\first{First flype-equivalent tangle diagrams.}{\epsfbox{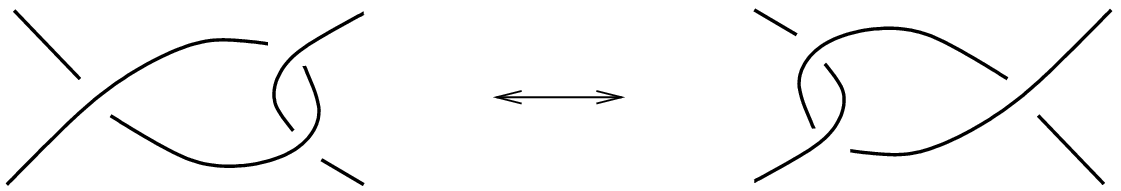}}

\vfill\eject

\vglue2cm
\centerline{\epsfxsize=25cm\epsfbox{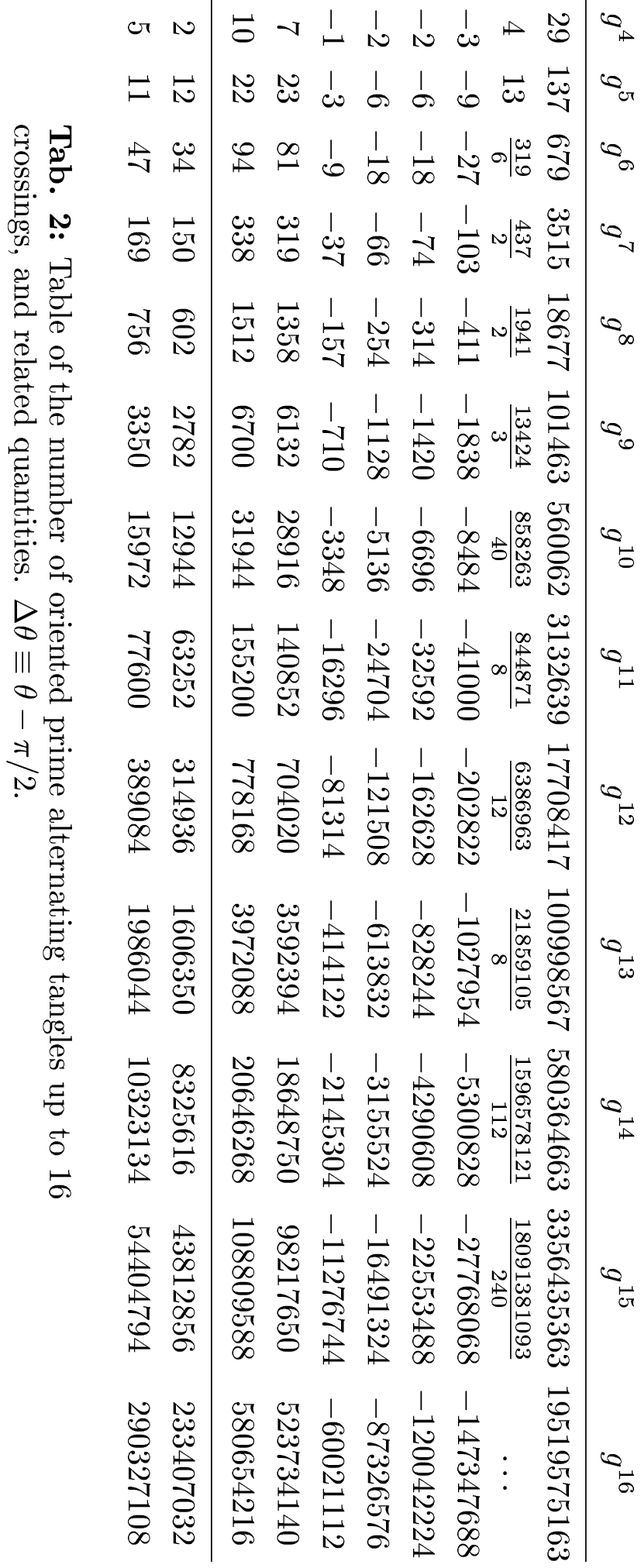}}

\vfill\eject

\subsec{Solution at criticality and large order behavior of the
correlation functions.}
In \PZJb, the model \sixv\ is analyzed in the critical regime, that
is when one encounters the first singularity of the free energy
(or of the correlation functions) by analytic continuation from
the gaussian model. This singularity -- or more precisely, the line
of singularities in the $(b_0,c_0)$ plane -- is what determines the
large order behavior of the correlation functions, and ultimately
of the functions $\Gamt$.  

We shall now summarize some of the relevant results from \PZJb.
If the parameter $\theta$ is fixed 
and the coupling constant $b_0$ is increased,
for any value of $\theta\in[0,\pi[$ there will be a critical value
$b_0^\star(\theta)$ for which the free energy becomes singular:
\eqn\crit{
b_0^\star(\theta)={1\over32}{\tan(\theta/2)\over\theta/2}
{1\over\cos^2(\theta/2)} \ .}
This forms a line of critical points,
describing what is known in physics as the
line of $c=1$ conformal field theories -- i.e. a 
free boson compactified on circles of varying radius -- coupled
to gravity, see for example \DFGZJ\ for a review and references.
In \PZJb, all correlation functions of the form 
$\lim_{N\to\infty}\left<{1\over N}\tr A^s\right>$ are calculated
on the critical line.
It is easy to see that the explicit expressions of 
$W_1=\lim_{N\to\infty}\left<{1\over N}\tr A\right>$
and
$W_2=\lim_{N\to\infty}\left<{1\over N}\tr A^2\right>$
allow to extract
the 2-point function and the free energy (still on the critical line);
one finds
\eqna\critexp
$$\eqalignno{
G^\star(\theta)&=8{\theta/2\over\tan(\theta/2)}
-{2\over 3}(\pi^2-\theta^2)&\critexp{\rm a}\cr
F^\star(\theta)&=
-4{\theta/2\over\tan(\theta/2)}
+{1\over 6}(\pi^2-\theta^2)\ . &\critexp{\rm b}\cr
}$$
By definition $F^\star(\theta)=F(b_0^\star(\theta)),\theta)$,
which means that
\eqn\critb{
{\d\over \d\theta} F^\star(\theta)={1\over 2 b_0^\star(\theta)}
G^\star(\theta)
{\d b_0^\star\over \d\theta}
+{\der\over\der\theta}F(b_0^\star(\theta),\theta)\ .}
This gives us access to
$H={\der\over\der\theta}F_{\left| b_0\ {\rm fixed}\right.}$
at criticality. One finds:
\eqn\critexpb{
H^\star(\theta)=-2\theta+{1\over2}(\pi^2-\theta^2)\tan(\theta/2)
-{1\over3}\pi^2{1\over\theta}+{1\over6}(\pi^2-\theta^2)\cot(\theta/2)\ .}
One must then use again the formulae \pertd--\perte{}\ and \dectwo{}\ 
to find the values of $\Gam_b$, $\Gam_c$, $D'_b$, $D'_c$
on the critical line as a function of $\theta$, and finally
solve \dectwob{} for $g$ and $\theta$. This is a set of
2 complicated coupled equations,\foot{In fact,
one of the two equations is quadratic in $g$, so that 
remains only one (transcendental) equation in $\theta$.}
and we can only solve them numerically.
We find the following numerical values:
\eqna\numval
$$\eqalignno{
\theta_c&=1.60780446\ldots&\numval{\rm a}\cr
1/g_c&=6.28329764\ldots&\numval{\rm b}\cr
}$$
The value \numval{\rm b}\ is non-universal, and only this explicit
calculation could give us access to it. It is related to the leading
behavior of the coefficients of the power series $\Gamt_b(g)$ and
$\Gamt_c(g)$, i.e.\ of the numbers
of oriented prime alternating tangles.
On the other hand, the subleading behavior is universal;
it is characteristic of $c=1$ conformal field theories coupled
to gravity, which exhibit zero ``string susceptibility''
with logarithmic corrections \refs{\PZJb,\IK}.
Therefore we can state that
if $\Gamt_b(g)=\sum \gamma_p g^p$, then
\eqn\asy{
\gamma_p{\buildrel p\to\infty\over\sim}
{\rm const}\ g_c^{-p} p^{-2} (\log p)^{-1}}
and similarly for $\Gamt_c$.
The constant $1/g_c=6.28329764\ldots$ is slightly less than the 
corresponding constant $16/(\pi(\pi-4)^2)=6.91167\ldots$ \PZJ\ for
oriented alternating link or tangle diagrams (without taking into account
the flype equivalence); and should also be compared to the numbers
obtained for $n=1$: $(101+\sqrt{21001})/40=6.147930\ldots$
and $27/4=6.75$ with and without the
flype equivalence, respectively. The fact
that the $n=1$ and $n=2$ results are fairly close
shows in particular that
the entropy generated by tangles with large numbers of connected
components is small.

As a final note, we can make some slightly conjectural statements
on the number of oriented prime alternating {\it links}. It is
clear that if we close
a tangle by pasting a simple crossing to its four external legs, we
obtain a link; if the tangle had $k$ closed loops, then
the number of connected components of the resulting link is simply
$k+2$ for a tangle of type 1 and $k+1$ for a tangle of type
2. Inversely, from a link with $p$ crossings
one can produce $4p$ tangles
by removing one of its vertices and fixing the circular permutation
of the four legs.
Even though we cannot use this fact to correctly count links
(one cannot make it a one-to-one correspondence), by assuming that
most links have low symmetry, we have the approximate relation
$f_p\approx \gamma_{p-1}/p$ satisfied by 
the number $f_p$ of links with $p$ crossings. Therefore 
we conjecture that 
\eqn\conjasy{
f_p{\buildrel p\to\infty\over\sim}
{\rm const}\ g_c^{-p} p^{-3} (\log p)^{-1}\ .}
\vskip1cm
\centerline{\bf Acknowledgements}
It is a pleasure to acknowledge stimulating discussions with
I.~Kostov. P.Z.-J. is supported in part by the DOE grant DE-FG02-96ER40559.
\footatend\vfill\supereject\immediate\closeout\rfile\writestoppt
\baselineskip=14pt\centerline{{\bf References}}\bigskip{\frenchspacing%
\parindent=20pt\escapechar=` \input refs.tmp\vfill\eject}\nonfrenchspacing

\bye